\documentclass[twocolumn,aps,prd,superscriptaddress,nofootinbib,amsmath,amssymb,floats,floatfix,showkeys,notitlepage,longbibliography]{revtex4-2}
\usepackage[dvips]{graphicx,color}
\usepackage{subfigure}
\usepackage{subcaption}
\usepackage{braket}
\usepackage{multirow}
\usepackage{times}
\usepackage{float}
\usepackage{tabularx}
\usepackage{makecell}
\usepackage{array}
\usepackage{mathrsfs}  
\usepackage{bm}
\usepackage{xfrac}
\usepackage{xcolor}
\usepackage[%
  colorlinks=true,
  urlcolor=blue,
  linkcolor=red,
  citecolor=blue
]{hyperref}
\usepackage{orcidlink}

\DeclareUnicodeCharacter{1E62}{\d{S}}
\newcommand\EMT{\mathrel{\stackrel{\makebox[0pt]{\mbox{\normalfont\tiny EM}}}{T}}}
\newcommand{\Lagr}{\mathcal{L}_m}

\newcommand{\hyper}{{}_2F_1}
\newcommand{\imp}{\text{b}}

\begin{document}
\title{GUP corrected Casimir Wormholes with Electric Charge in \texorpdfstring{$f(R, \Lagr)$}{} Gravity}

\author{Mohammed Muzakkir Rizwan \orcidlink{0009-0008-6375-9045}}
\email{mohammed.muzu.riz@gmail.com}
\affiliation{Department of Mathematics, Birla Institute of Technology and
Science-Pilani,\\ Hyderabad Campus, Hyderabad-500078, India.}

\author{Zinnat Hassan \orcidlink{0000-0002-3223-4085}}
\email{zinnathassan980@gmail.com}
\affiliation{Department of Mathematics, Birla Institute of Technology and
Science-Pilani,\\ Hyderabad Campus, Hyderabad-500078, India.}

\author{P.K. Sahoo \orcidlink{0000-0003-2130-8832}}
\email{pksahoo@hyderabad.bits-pilani.ac.in}
\affiliation{Department of Mathematics, Birla Institute of Technology and
Science-Pilani,\\ Hyderabad Campus, Hyderabad-500078, India.}
\date{\today}

\begin{abstract}
In this letter, we study and investigate the effects of the Generalised Uncertainty Principle (GUP) and electric charge on Casimir wormhole geometry in the Curvature-Lagrangian coupled $f(R,\Lagr)$ gravity. The functional form of the considered model is $f(R,\Lagr)=\frac{R}{2\kappa}+\Lagr^{\,\,\alpha}$, corresponding to it the analytic shape function is found. For our analysis, we study the wormhole spacetimes for three particular models for the redshift function. We observe that the null energy condition is violated despite a positive contribution from the electromagnetic energy density. We also note that electric charges, GUP effects, and higher model parameter values increase the throat length. Further, we have studied the deflection of light using the Gauss-Bonnet theorem, emphasizing the contribution from GUP by including higher-order terms.
\end{abstract}

\maketitle
\section{Introduction}
Wormholes are hypothetical bridges connecting the spacetime manifold of different universes, branes, or even two points within the same universe. The first known theoretical formulation for a wormhole was by Flamm \cite{flamm1916beitrage}. Thereafter, an extended study was done by Einstein and Rosen \cite{einstein1935particle}, which brought attention to many of what was called the Einstein-Rosen bridge. The Einstein-Rosen bridge was a wormhole construction using the maximally stretched Schwarzschild metric, connecting two universes. Due to the event horizons, any object or person entering cannot exit the wormhole and will fall to the singularity. This is despite the fact that the wormhole throat is highly unstable for the slightest of perturbations, which would collapse for even light to pass through it. Later in 1988, Morris and Thorne \cite{morris1988wormholes} propounded the idea of traversable wormholes that would allow an external object or traveler to pass through them, allowing interstellar travel. This, however, is only possible to achieve through exotic matter for stabilizing the wormhole spacetime geometries. This leads to the violation of the Null Energy Condition (NEC), which classically is not possible to achieve. Unless we can intrinsically alter the formulation of GR to allow negative energy density or negative pressures. Interestingly, the Casimir effect provides a possibility to produce negative energy density in regions of space, and this can be used to stabilize traversable wormholes.\\
\indent The Casimir effect proves to be a viable and practical source of exotic matter. It was theoretically proposed by H. Casimir in 1948 \cite{casimir1948attraction} and independently studied by \cite{dzyaloshinskii1961general}, later experimentally verified in the Philips laboratory \cite{black1960measurements}. However, it was only in recent years that more reliable experimental investigations were employed to confirm this phenomenon \cite{lamoreaux1997demonstration,bressi2002measurement}. Moreover, the classical electric and magnetic fields can contribute to the stabilization of the wormhole. In fact, charged wormholes can be thought of as an extension of the Morris-Thorne wormholes \cite{kuhfittig2011feasibility}, where the presence of charges plays the role of additional matter content to the existing exotic matter \cite{kim2001exact}. However, certain wormhole geometries can sustain without exotic matter by virtue of the type of elastic potential associated with the spacetime or geometry-matter coupling in the background, which is not the case for GR. Here are a few examples of such cases in $f(R)$ gravity \cite{capozziello2021traversable,de2021reconstructing,Capozziello:2022zoz,de2021testing}. A formal solution for the possibility of Casimir wormholes in GR was put forth by R. Garattini \cite{garattini2019casimir}, followed by an extended study on the same but with the presence of static electric charges on the wormhole's throat \cite{garattini2023effects}. Later, Jusufi et al. \cite{jusufi2020traversable} studied the effect of minimal length correction on the Casimir energy inspired by the GUP in the context of GR, while Samart et al. \cite{samart2022charged} used this as a bedrock for their analysis by considering the GUP Casimir wormhole possessing an electric charge. The stability of GUP Casimir wormholes in the Rastall-Rainbow gravity is examined using the Herrera cracking technique in \cite{battista2024generalized}. Also, Casimir wormholes \cite{ZinnatGUP1}, including GUP Casimir wormhole \cite{ZinnatGUP2}, have been discussed in $f(Q)$ gravity.
Further, in the realm of Curvature-Lagrangian coupled $f(R,\Lagr)$ gravity, the possibility of traversable Casimir wormholes has been studied in \cite{khatri2024casimir}.\\
\indent In recent years, \cite{harko2010f} proposed $f(R,\Lagr)$ gravity, a generalization of curvature-matter coupling theories where $f(R,\Lagr)$ is a function of the Ricci scalar `$R$' and the Lagrangian density `$\Lagr$'. The covariant divergence of the energy-momentum tensor is non-zero due to extra forces from these couplings, causing deviations from the geodesic equation in GR. The effective energy-momentum tensor, however, is conserved, which takes into account the additional force. Cosmological models with $f(R,\Lagr)$ gravity violate the equivalence principle and are constrained by solar system tests \cite{faraoni2004scalar,bertolami2008general}. Recent studies using this framework include \cite{lobato2021neutron,harko2022palatini,harko2015cosmic}, exploring cosmological and astrophysical phenomena. Wormhole solutions have also been studied within this theory, such as those using various equations of state (EoS) \cite{solanki2023wormhole} and non-commutative geometry with Gaussian and Lorentzian distributions \cite{kavya2023possible}. Recently, static and conformally symmetric traversable wormhole solutions supported by phantom dark energy were found in \cite{taser2024conformal}.\\
\indent The main aim of this letter is to study the effects of GUP on the Casimir wormhole, which possesses a static net electric charge on the throat. 
The article is organized as follows: the theoretical premise for the $f(R,\Lagr)$ theory, traversable wormholes, and modified Einstein-Maxwell equations are derived in Sec.\ref{sec: WH theory}. The GUP Casimir effect is briefly explained in Sec.\ref{sec: CE GUP}. In Sec.\ref{sec: WH sol}, the analytic solution for the wormhole geometry is discussed using three redshift functions. The Gauss-Bonnet Theorem is applied to find the deflection of light rays approaching the wormhole in Sec.\ref{sec: Lensing}. Finally, Sec.\ref{sec: conclusions} concludes the article.
\section{Charged Wormholes in \texorpdfstring{$f(R, \Lagr)$}{} formalism}\label{sec: WH theory}
The Einstein-Hilbert action describing the dynamics of $f(R, \Lagr)$ modified gravity and electromagnetism is
\begin{equation}\label{action}
    \mathcal{S}=\int \Bigr[f(R, \Lagr) - \frac{1}{4\mu_0}F_{\mu\nu}F^{\mu\nu}\Bigl] \sqrt{-g}d^4x,
\end{equation}
where $F_{\mu\nu}$ is the electromagnetic tensor, $g$ is the determinant of the metric, $f(R, \Lagr)$ is an arbitrary function involving the Ricci Scalar `$R$' and matter Lagrangian density `$\Lagr$'. The Ricci curvature is given by the contraction of the Ricci tensor $R_{\mu\nu}$ as
\begin{equation}\label{ricci scalar}
    R=g^{\mu\nu}R_{\mu\nu},
\end{equation}
where
\begin{equation}\label{ricci tensor}
    R_{\mu\nu}=\partial_\lambda\Gamma^\lambda_{\mu\nu}-\partial_\mu\Gamma^\lambda_{\lambda\nu}+\Gamma^\lambda_{\mu\nu}\Gamma^\sigma_{\sigma\lambda}-\Gamma^\lambda_{\nu\sigma}\Gamma^\sigma_{\mu\lambda}.
\end{equation}
$\Gamma^l_{ij}$ are the components of the Levi-Civita connection. The energy-momentum tensor for the matter fields is obtained by
\begin{align}\label{EM tensor}
    T_{\mu\nu}=-\frac{2}{\sqrt{-g}} \frac{\delta\left(\sqrt{-g} \Lagr\right)}{\delta g^{\mu \nu}}
    =g_{\mu\nu}\Lagr-2\frac{\partial\Lagr}{\partial g^{\mu\nu}}.
\end{align}
The energy-momentum tensor associated with the electrodynamics is
\begin{equation}\label{EMT Tensor}
    \EMT_{\mu\nu}=\frac{1}{\mu_0}\left(g_{\nu\alpha}F_{\mu\gamma}F^{\alpha\gamma}-\frac{1}{4}g_{\mu\nu}F_{\alpha\beta}F^{\alpha\beta}\right).
\end{equation}
On varying the action \eqref{action} with respect to the metric tensor, we have the following equations of motion at hand:
\begin{equation}\label{eqs of motion}
    f_RR_{\mu\nu}+(g_{\mu\nu}\square-\nabla_\mu\nabla_\nu)f_R-\frac{1}{2}(f-f_{\Lagr}\Lagr)g_{\mu\nu}=\frac{1}{2}\mathcal{T}^{eff}_{\mu\nu},
\end{equation}
where $f_R\equiv\frac{\partial f}{\partial R}$, $f_{\Lagr}\equiv\frac{\partial f}{\partial\Lagr}$ and $\mathcal{T}^{eff}_{\mu\nu}=f_{\Lagr}T_{\mu\nu}+\EMT_{\mu\nu}$. On contracting the field equation \eqref{eqs of motion}, we can acquire the relation between the Ricci scalar $R$, Lagrangian density term $\Lagr$ and the energy-momentum scalar $\mathcal{T}^{eff}$
\begin{equation}\label{contracted EFE}
    f_RR+3\square f_R-2(f-f_{\Lagr}\Lagr)=\frac{1}{2}\mathcal{T}^{eff},
\end{equation}
where $\square\Psi=\frac{1}{\sqrt{-g}}\partial_\mu(\sqrt{-g}g^{\mu\nu}\partial_\nu\Psi)$ for any scalar function $\Psi$. Furthermore, on taking the divergence of \eqref{eqs of motion} we obtain the following:
\begin{equation}\label{EM tensor consv}
    \nabla^\mu T_{\mu\nu}=2\nabla^\mu ln(f_{\Lagr})\frac{\partial\Lagr}{\partial g^{\mu\nu}}.
\end{equation}
Now, we consider the Morris-Thorne metric \cite{morris1988wormholes} that describes a static and spherically symmetric wormhole, defined as
\begin{equation}\label{metric}
    ds^2=-e^{2\Phi(r)}dt^2+\frac{dr^2}{1-b(r)/r}+r^2(d\theta^2+\text{sin}^2\theta d\phi^2),
\end{equation}
where the redshift function $\Phi(r)$ relates to gravitational redshift. It must be finite everywhere to avoid event horizons. While $b(r)$ is the shape function that determines the geometry of the wormhole and must satisfy the following conditions: $(i)$ $b(r_0)=r_0$, $(ii)$ $b'(r)<1$ at $r=r_0$, $(iii)$ $b(r)-rb'(r)/b^2(r)>0$, $(iv)$ $b(r)<r$ for $r>r_0$, and $(v)$ $b(r)/r \to 0$ as $r \to \infty$. Condition $(iv)$ implies $(iii)$, which reduces to $(ii)$ at the throat.
The embedding of the wormhole can be expressed by fixing $\theta=\pi/2$ and $t=\text{constant}$ which yields the metric
\begin{equation}
    ds^2=\frac{dr^2}{1-b(r)/r}+r^2d\phi^2.
\end{equation}
This can be embedded in a three-dimensional cylindrically symmetric Euclidean space, which has the metric: $ds^2=dz^2+dr^2+r^2d\phi^2$. On comparing both of these metrics, we can find the embedding surface $z(r)$
\begin{equation}
    \frac{dz}{dr}=\pm\left(\frac{r}{b(r)}-1\right)^{-1/2}.
\end{equation}
For the metric \eqref{metric}, the Ricci scalar takes the value
\begin{equation}\label{R scalar val}
    R=\frac{2b'}{r^2}-2\left(1-\frac{b}{r}\right)\Biggl[\Phi''+\Phi'^2+\frac{2\Phi'}{r}-\frac{(rb'-b)\Phi'}{2r(r-b)}\Biggr],
\end{equation}
and the D'Alembertian acting on a scalar function of the radial coordinate $F=F(r)$ in such a spacetime is evaluated as
\begin{equation}\label{box oprtr val}
    \square F=\left(1-\frac{b}{r}\right)\Biggl\{\Biggl[\Phi'+\frac{2}{r}-\frac{rb'-b}{2r(r-b)}\Biggr]F'+F''\Biggr\}.
\end{equation}
In this letter, we shall consider the anisotropic matter fluid, whose energy-momentum tensor looks like 
\begin{equation}\label{emtensor}
    T^\mu_\nu=(\rho+p_t)u^\mu u_\nu+p_t\delta_\nu^\mu+(p_r-p_t)v^\mu v_\nu,
\end{equation}
where $u^\mu=e^{-\Phi(r)}\delta^\mu_t$ is the four-velocity, and $v^\mu=\sqrt{1-b(r)/r}\delta^\mu_r$ is the unit space-like vector in the radial direction. The components $\rho,\ p_r$ and $p_t$ respectively denote the energy density, radial pressure, and tangential pressure, which make up the diagonal matrix $T^\mu_\nu=\text{diag}(-\rho, p_r, p_t, p_t)$.\\
\indent We are concerned with static charges present at the throat of the wormhole, which emanate a spherically symmetric electric field in the radial direction at large distances as $\mathbf{E}=E_r(r)\hat{r}=\frac{Q}{4\pi\epsilon_0r^2}\hat{r}$, where `$Q$' is the total charge present at the throat. It is to be noted that, at considerable distances, the charge distribution effectively acts as a point source, hence we go by this assumption. Naturally, the magnetic field is zero. With all these considerations, the electromagnetic (EM) energy-momentum tensor becomes:
\begin{equation}\label{emt tensor}
    \EMT^{\;\mu}_\nu=u\ \text{diag}(-1,-1,1,1),
\end{equation}
where `$u$' is the electromagnetic energy density $u=\frac{1}{2}\epsilon_0{|\mathbf{E}|}^2=\frac{Q^2}{32\pi^2\epsilon_0r^4}$. Considering the above information, our effective energy-momentum tensor is:
\begin{equation}\label{eff em tensor}
    {\mathcal{T}^{eff}}^\mu_\nu=\text{diag}\left(-\rho^{eff},\ p_r^{eff},\ p_t^{eff},\ p_t^{eff}\right),
\end{equation}
where $\rho^{eff}=f_{\Lagr}\rho+u,\ p_r^{eff}=f_{\Lagr}p_r-u,\ p_t^{eff}=f_{\Lagr}p_t+u$. Thus by inserting the metric \eqref{metric}, and the effective energy-momentum tensor \eqref{eff em tensor} into \eqref{eqs of motion}, we obtain the following field equations:
\begin{multline}\label{EFE1}
    \left(1-\frac{b}{r}\right)\Biggl\{\Biggl[\Phi''+\Phi'^2+\frac{2\Phi}{r}-\frac{rb'-b}{2r(r-b)}\Phi'\Biggr]f_R-\Biggl[\Phi'+\frac{2}{r}\\-\frac{rb'-b}{2r(r-b)}\Biggr]f_R'-f_R''\Biggr\}+\frac{1}{2}(f-f_{\Lagr}\Lagr)=\frac{\rho^{eff}}{2},
\end{multline}
\begin{multline}\label{EFE2}
    \left(1-\frac{b}{r}\right)\Biggl\{\Biggl[-\Phi''-\Phi'^2+\frac{rb'-b}{2r(r-b)}\left(\Phi'+\frac{2}{r}\right)\Biggr]f_R+\Biggl[\Phi'\\+\frac{2}{r}-\frac{rb'-b}{2r(r-b)}\Biggr]f_R'\Biggr\}-\frac{1}{2}(f-f_{\Lagr}\Lagr)=\frac{p_r^{eff}}{2},
\end{multline}
\begin{multline}\label{EFE3}
    \left(1-\frac{b}{r}\right)\Biggl\{\Biggl[-\frac{\Phi'}{r}+\frac{rb'+b}{2r^2(r-b)}\Biggr]f_R+\Biggl[\Phi'+\frac{2}{r}\\-\frac{rb'-b}{2r(r-b)}\Biggr]f_R'+f_R''\Biggr\}-\frac{1}{2}(f-f_{\Lagr}\Lagr)=\frac{p_t^{eff}}{2}.
\end{multline}
Here the $'$ denotes the derivative with respect to $r$.
\section{A brief description of the Casimir Effect and The GUP}\label{sec: CE GUP}
The theoretical existence of a traversable wormhole can only manifest into practicality when the energy conditions are violated at the wormhole's throat. In other words, the possibility of exotic matter's existence is paramount for a sustainable wormhole geometry. Although the exotic matter was hypothesized as a tool to bypass the energy constraints mathematically, our current knowledge of Quantum Field Theory (QFT) indicates that vacuum fields can violate the NEC in certain setups. The Casimir effect is the phenomenon in QFT where, when two uncharged conducting plane plates are held parallel together at an extremely close distance in a vacuum, an apparent attractive force manifests between them. The metal plates act as boundary conditions to the quantum electrodynamics, which result in a negative zero-point energy.\\
The renormalized energy between the conducting plates is mathematically expressed as
\begin{equation}\label{casimirenergy}
    E(a)=-\frac{\hbar c\pi^2A}{720a^3},
\end{equation}
where `$a$' is the separation between the parallel plates and `$A$' is their surface area. One can derive the expression \eqref{casimirenergy} by summing over the normal modes of the vacuum field and adequately regularizing the sum \cite{zee2010quantum,padmanabhan2016quantum}. Consequently, the energy density can be found by dividing (\ref{casimirenergy}) with the volume between the plates $V=a\times A$
\begin{equation}\label{casimirdensity}
    \rho_c=\frac{E}{aA}=-\frac{\hbar c\pi^2}{720a^4}.
\end{equation}
The pressure on the surface of the plates is
\begin{equation}
    p_c=\frac{F_c}{A}=-\frac{1}{A}\frac{dE}{da}=-3\frac{\hbar c\pi^2}{720a^4}.
\end{equation}

Thus, obeying the linear EoS, $\rho_c=\omega p_c$, with $\omega=3$. Throughout the literature on quantum gravity, the minimal length concept from quantum mechanics is consolidated, limiting the minimal measurable length beyond which spacetime cannot be resolved further. As a consequence, it is pertinent to modify the definitions of linear momentum as well as the quantum commutation relations which implies modified dispersion relations, e.g., gravity’s rainbow \cite{magueijo2004gravity}; certain cosmological \cite{chatrabhuti2016starobinsky,channuie2019deformed,hendi2016nonsingular} and astrophysical inferences \cite{hendi2017charged,feng2017thermodynamic,hendi2018ads,panahiyan2019ads4,dehghani2018thermodynamics}. This is not as non-trivial as it seems at first glance, as such a scale arises naturally in theories of gravity usually in the form of an effective minimal uncertainty in positions $\Delta x_0>0$. Such as, in the case of string theory, it is impossible to exceed the spatial resolution beyond the characteristic length of the strings. This necessitates correction to the uncertainty principle, which in one dimension looks like\\
\begin{equation}
    \Delta x\Delta p\geq\frac{\hbar}{2}\Bigl[1+\lambda(\Delta p)^2+\eta\Bigr],\;\;\;\;\;\lambda,\eta>0
\end{equation}
where the finite minimal length uncertainty appears as $\Delta x_0=\hbar\sqrt{\lambda}$, in terms of the minimum length parameter $\lambda$. Consequently, this new uncertainty relation implies the following modification to the usual Heisenberg commutation relation as
\begin{equation}\label{gup 1d}
    [\hat{x},\ \hat{p}]=i\hbar\left(1+\lambda{\hat{p}}^2+\cdots\right).
\end{equation}
We also note that position and momentum are no longer conjugate variables in the usual sense; hence, in these theories, the position eigenstates can no longer be interpreted as actual physical positions. However, we can still talk about the positions as the states projected onto the set of maximally localized states, called the ``quasi-position representation". These maximally localized states $|\psi^{ML}_x\rangle$ minimize the uncertainty in position $(\Delta x)_{|\psi^{ML}_x\rangle}=\Delta x_0$ and are centered around some average position $\langle\psi^{ML}_x|\hat{x}|\psi^{ML}_x\rangle=x$. The commutation relation in \eqref{gup 1d} can be generalized for $n$ spatial dimensions, that leads to the GUP that provides
a minimal uncertainty in position, as
\begin{equation}\label{GCR}
    [\hat{x}_i,\ \hat{p}_j]=i\hbar\Bigl[f\left(\hat{p}^2\right)\delta_{ij}+g\left(\hat{p}^2\right)\hat{p}_i\hat{p}_j\Bigr],
\end{equation}
for $i,j=1,2,\ldots,n$. $f(\hat{p}^2)$ and $g(\hat{p}^2)$ are generic functions. These functions are not completely arbitrary, and relations between them can obtained by imposing the translational and rotational invariance on the generalized commutation relation \eqref{GCR}. In literature, the popular GUP corrections are the Kempf, Mangano, and Mann (KMM) approach, the Detournay, Gabriel, and Spindel (DGS) approach, and the so-called Model II. The choice of generic functions for these methods is as follows: \\
\indent Model I (KMM \& DGS):
\begin{equation}
    f(\hat{p}^2)=\frac{\lambda \hat{p}^2}{\sqrt{1+2\lambda\hat{p}^2}-1},\;\;\;g(\hat{p}^2)=\lambda,
\end{equation}
\indent Model II:
\begin{equation}
    f(\hat{p}^2)=1+\lambda \hat{p}^
    2,\;\;\;g(\hat{p}^2)=0.
\end{equation}
 Although the KMM and DGS assume the same choice of generating functions, these approaches utilize different choices of the measure function $\Omega(p)$ detailed in \cite{frassino2012casimir}. Thus, by incorporating the concept of minimal length and the GUP, the authors of \cite{frassino2012casimir} have derived the Hamiltonian and corrections to the Casimir Energy up to the first order of the minimal length parameter $\lambda$. The corrected form of Casimir Energy is
\begin{equation}\label{GUPenergy}
    E_i(a)=-\frac{\hbar c\pi^2}{720}\frac{A}{a^3}\Biggr[1+\Lambda_i\Biggr(\frac{\hbar^2\lambda}{a^2}\Biggr)\Biggr],
\end{equation}
where $\Lambda_i$ is a constant for $i=\text{KMM},\ \text{DGS},\ \text{Model II}$ which corresponds to different models
\begin{equation*}
    \Lambda_{KMM}=\pi^2\Biggl(\frac{28+3\sqrt{10}}{14}\Biggl),\quad\Lambda_{DGS}=4\pi^2\Biggl(\frac{3+\pi^2}{21}\Biggl),
\end{equation*}
$$\Lambda_{Model\ II}=\frac{2\pi^2}{3}.$$
Hence, the expression for the Casimir pressure on the plates is,
\begin{equation}
    P(a)=-\frac{1}{A}\frac{dE}{da}=-\frac{3\hbar c\pi^2}{720}\frac{1}{a^4}\Biggr[1+\frac{5}{3}\Lambda_i\Biggr(\frac{\hbar^2\lambda}{a^2}\Biggr)\Biggr].
\end{equation}
Using the EoS relation for the Casimir Effect as established before, i.e., for $\omega=3$, we get the GUP corrected Casimir energy density
\begin{equation}\label{GUPdensity}
    \rho(a)=-\frac{\hbar c\pi^2}{720a^4}\Biggr[1+\frac{5}{3}\Lambda_i\Biggr(\frac{\hbar^2\lambda}{a^2}\Biggr)\Biggr].
\end{equation}
\indent In the methodology that follows, we promote the plate separation distance to the radial coordinate, which is justifiable as we are dealing with lengths in the order of Planck length. We also define the term $r_i^2={r_i(\lambda)}^2=\frac{5}{3}\hbar^2\Lambda_i\lambda$ that has the dimensions of $[length]^2$. Additionally, we will make use of the following expressions just as they were in \cite{garattini2023effects}
\begin{equation}
    r_1^2=\frac{\pi^3l_p^2}{90},\;\;\;\;r_2^2=\frac{G\,Q^2}{4\pi\epsilon_0c^4},
\end{equation}
where $l_p$ is the Planck length. Throughout this letter, we shall retain the expressions in the SI convention, unless mentioned otherwise, or for plotting, in which case we employ the following natural units $G=\hbar=c=1$ and $\epsilon_0=1/4\pi$.
\section{Charged Wormhole solutions with GUP corrected Casimir source}\label{sec: WH sol}
Our consideration for the model function is inspired by the generic $f(R,\Lagr)$ function; specifically, $f(R,\Lagr)=f_1(R)+f_2(R)G(\Lagr)$ that represents an arbitrary curvature-matter coupling \cite{harko2014generalized}. We presume the following non-linear function used in \cite{solanki2023wormhole,khatri2024casimir}
\begin{equation}\label{model 1}
    f(R,\Lagr)=\frac{R}{2\kappa}+\Lagr^{\,\,\alpha},
\end{equation}
where $\alpha$ is a free parameter and $\kappa=8\pi G/c^4$ is Einstein's gravitational constant. It is trivial that by setting $\alpha=1$ we retrieve the premise of GR. We have chosen the matter Lagrangian to be $\Lagr=\rho$.
Thus our field equations (\ref{EFE1}-\ref{EFE3}) become
\begin{equation}\label{rhoM1}
    \rho^{eff}=\alpha\rho^\alpha+u=-(\alpha-1)\rho^\alpha+\frac{b'}{\kappa r^2},
\end{equation}
\begin{multline}\label{prM1}
    p_r^{eff}=\alpha\rho^{\alpha-1}p_r-u=(\alpha-1)\rho^\alpha+\frac{1}{\kappa}\Biggl[-\frac{b}{r^3}+\frac{2}{r}\left(1-\frac{b}{r}\right)\Phi'\Biggr],
\end{multline}
\begin{multline}\label{ptM1}
    p_t^{eff}=\alpha\rho^{\alpha-1}p_t+u=(\alpha-1)\rho^\alpha+\frac{1}{\kappa}\Biggl\{\left(1-\frac{b}{r}\right)\Biggl[\Phi''\\+\Phi'\left(\Phi'+\frac{1}{r}\right)\Biggr]-\frac{rb'-b}{2r^2}\left(\Phi'+\frac{1}{r}\right)\Biggr\}.
\end{multline}
\begin{figure*}
\centering
    \subfigure[$b(r)$]{\includegraphics[scale=0.35]{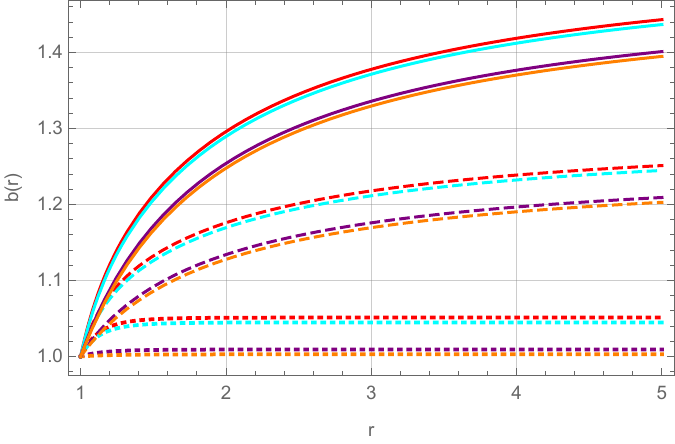}}
    \subfigure[$b'(r)$]{\includegraphics[scale=0.35]{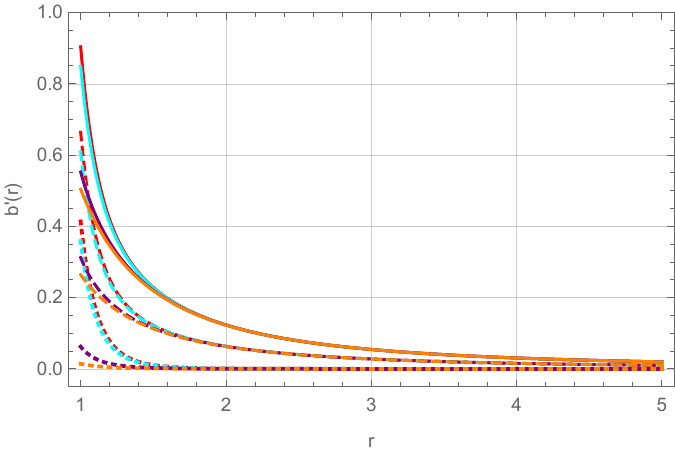}}
    \subfigure[$b(r)/r$]{\includegraphics[scale=0.35]{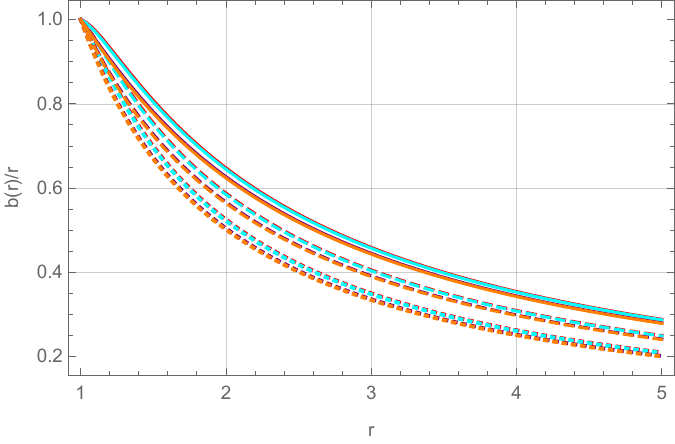}\label{subfig: brM1}}
    \subfigure[$\rho^{eff}(r)$]{\includegraphics[scale=0.35]{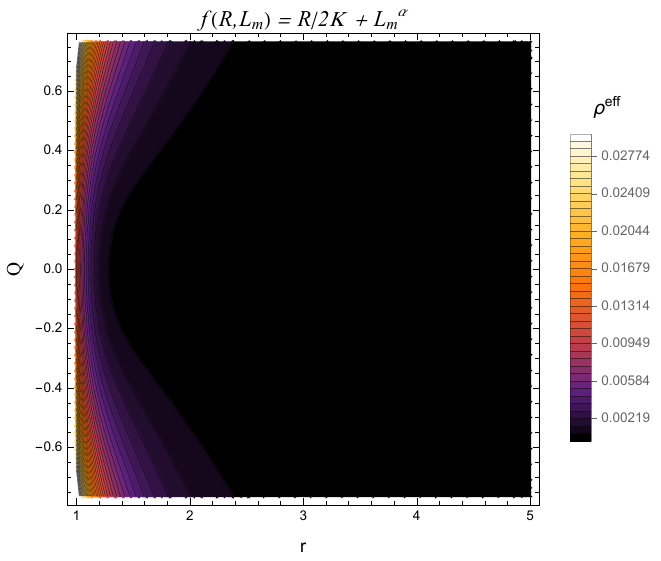}\label{subfig: eff-rhoM1}}
\caption{Plots depicted the shape function solution and its properties. KMM is represented by red, DGS by cyan, Model II by purple, and non-GUP by orange, while dotted plots are for uncharged $Q=0$, dashed for $Q=0.5$, and solid curves for $Q=0.7$. Plot-(d) portrays the effective energy density of the matter source for the KMM case with $\alpha=2$, $\lambda=0.1$, and $r_0=1$.}
    \label{fig:Shape function M1}
\end{figure*}
\begin{figure*}
\centering
    \subfigure[]{\includegraphics[scale=0.4]{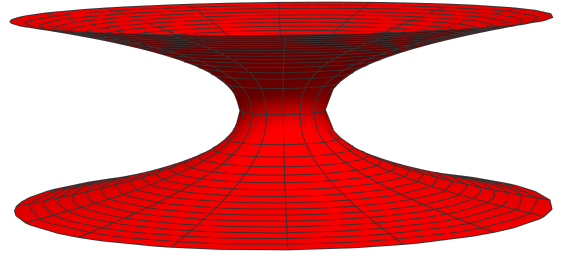}}\qquad
    \subfigure[]{\includegraphics[scale=0.4]{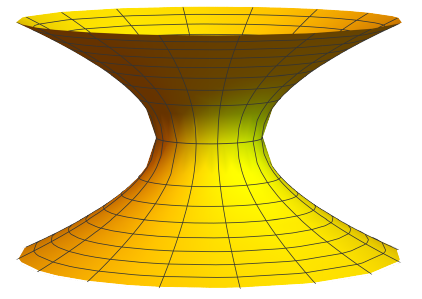}}\qquad
    \subfigure[]{\includegraphics[scale=0.4]{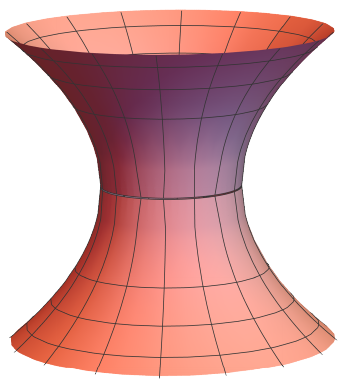}}\qquad
    \subfigure[]{\includegraphics[scale=0.4]{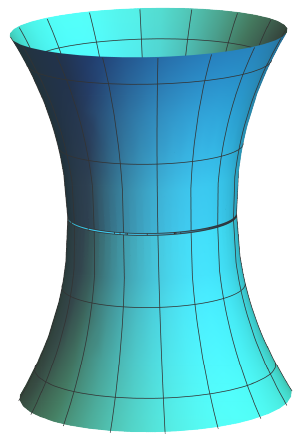}}
\caption{The embedding diagrams for Casimir wormholes with the backgrounds (a) $\alpha=1$, $Q=0$, (b) $\alpha=1$, $Q=0.765$, (c) $\alpha=2$, $Q=0$ and (d) $\alpha=2$, $Q=0.765$. KMM case is considered here, with $\lambda=0.1$ and $r_0=1$.}
    \label{fig:embedding}
\end{figure*}
From \eqref{GUPdensity} and \eqref{rhoM1}, we find the shape function as
\begin{equation}\label{shfuncM1}
    b(r)=r_0+\mathcal{M}(r;\alpha)-\mathcal{M}(r_0;\alpha)-r_2^2\left(\frac{1}{r}-\frac{1}{r_0}\right),
\end{equation}
where we have defined the following function $\mathcal{M}(r;\alpha)=\frac{2\alpha-1}{3-4\alpha}\frac{(-r_1^2)^\alpha}{\kappa^{\alpha-1}}\Bigl[r^{3-4\alpha}\hyper(-\alpha,\ 2\alpha-\frac{3}{2},\ 2\alpha-\frac{1}{2},-\frac{r_i^2}{r^2})\Bigr]$, that encodes the contribution from Casimir Energy and GUP effects. It becomes evident that the asymptotic flatness condition for the above shape function holds only if $\alpha>1/2$. Additionally, the presence of ${(-r_1^2)}^\alpha$ term conveys that we can only use integer values for the model parameter $\alpha$ for $b(r)$ to be a real-valued function. Furthermore, the flare-out condition imposes a bound on what values $r_2$ can have, which is given by
\begin{equation}\label{37}
    |r_2|<r_0\sqrt{1-(2\alpha-1)\kappa{r_0}^2{\rho_0}^\alpha},
\end{equation}
where $\rho_0=\rho(r_0)$. For our analysis, we have considered $\alpha=2$, $r_0=1$, and the correction parameter $\lambda=0.1$. For the KMM case, the charge is bounded roughly as $|Q|<0.7656$ in the aforementioned choice of natural units.
We observe in Fig.\ref{fig:Shape function M1} that our shape function solution is coherent with the assumptions for the Morris-Thorne metric. We observe that the flare-out condition is satisfied under the asymptotic background.\\
The effective energy density of the system, which takes into account the electrostatic effects and the minimal coupling of the model, is
\begin{equation}\label{rhoeffM1}
    \rho^{eff}(r)=\frac{\alpha{(-r_1^2)}^\alpha{(r^2+r_i^2)}^\alpha r^4+\kappa^{\alpha-1}r_2^2r^{6\alpha}}{\kappa^\alpha r^{6\alpha+4}}.
\end{equation}
From Fig.\ref{subfig: eff-rhoM1}, the effective energy density of the source matter is positive for $\alpha=2$. Additionally, the embedding diagrams in Fig.\ref{fig:embedding} display the stark difference between the wormhole geometry in the case of GR and modified gravity, with or without electric charge, in particular, lengthening of the region near the throat. Using the shape function \eqref{shfuncM1} and the linear barotropic EoS $p_r^{eff}=\omega\rho^{eff}$ in the field equation \eqref{prM1}, we find that the redshift function $\Phi(r)$ does not have an analytic solution. Thus, we consider the same three models for the redshift function in \cite{samart2022charged} to keep up the spirit of this paper as a succession to theirs. We now study the energy conditions entailed by such tidal effects.
\subsubsection{\texorpdfstring{$\Phi(r)=\text{constant}$}{}}
A constant redshift function can be treated as $\Phi(r)=0$ without loss of generality as $e^{\Phi(r)}$, in such case, can be absorbed in the $dt^2$ term of the metric. The radial and tangential components of pressure for spacetime with zero tidal forces is
\begin{equation}
    p_r^{eff}(r)=\frac{\mathcal{H}_1r^4-\kappa^{\alpha-1}r^{6\alpha}\mathcal{E}_1}{\kappa^\alpha r^{6\alpha+4}r_0},
\end{equation}
\begin{equation}
    p_t^{eff}(r)=\frac{2\mathcal{H}_1r^4+\kappa^{\alpha-1}r^{6\alpha}\mathcal{E}_2}{2\kappa^\alpha r^{6\alpha+4}r_0},
\end{equation}
\begin{figure*}
    \includegraphics[scale=0.4]{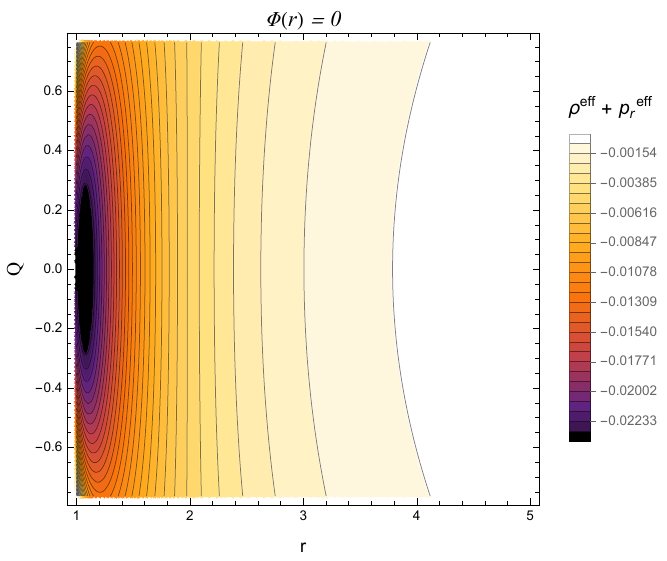}\qquad
    \includegraphics[scale=0.4]{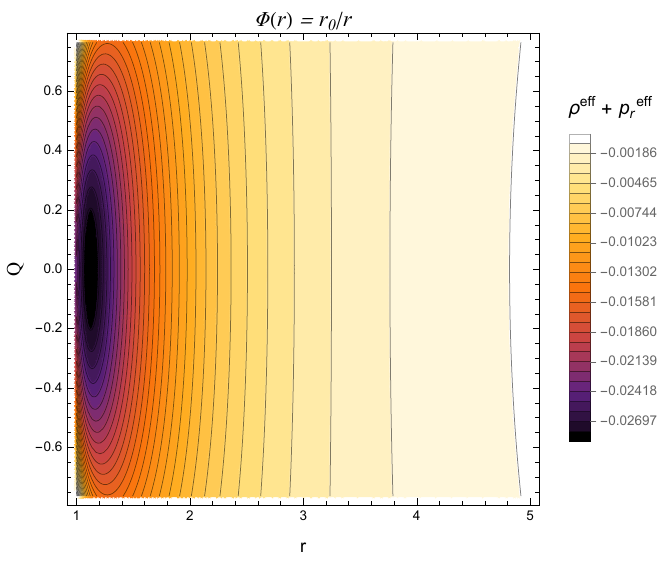}\qquad
    \includegraphics[scale=0.4]{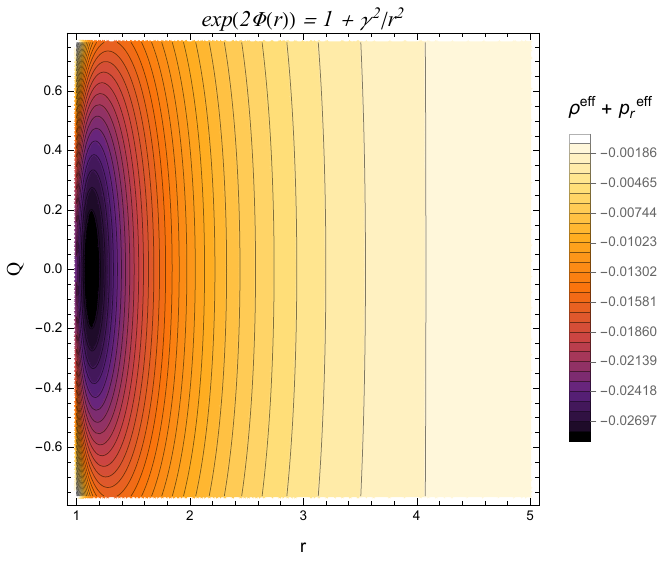}
    \caption{Contour plots for the radial part of Null Energy Condition: $\rho^{eff}+p_r^{eff}$ by varying the electric charge as $-0.765\leq Q\leq 0.765$ for the KMM case, $\lambda=0.1$, $\alpha=2$, $\gamma=3$ and $r_0=1$ in natural units.}
    \label{fig:NECrM1} 
\end{figure*}
with the following parameters defined as
\begin{equation}
    \mathcal{H}_1=(\alpha-1){(-r_1^2)}^\alpha{(r^2+r_i^2)}^\alpha r_0,
\end{equation}
\begin{equation}
    \mathcal{E}_1=r_0^2r+r_0r\mathcal{M}(r;\alpha)-r_0r\mathcal{M}(r_0;\alpha)-r_2^2(r_0-r),
\end{equation}
\begin{equation}
    \mathcal{E}_2=\mathcal{E}_1-r_0r^2\mathcal{M'}(r;\alpha)-r_2^2r_0,
\end{equation}
where $\mathcal{M'}(r;\alpha)=\frac{d}{dr}\mathcal{M}(r;\alpha)$.
\subsubsection{\texorpdfstring{$\Phi(r)=r_0/r$}{}}
This is a typical example of a non-constant redshift function. The pressure components for this geometry are
\begin{equation}
    p_r^{eff}(r)=\frac{\mathcal{H}_1r^5-\kappa^{\alpha-1}r^{6\alpha}\mathcal{G}_1}{\kappa^\alpha r^{6\alpha+5}r_0},
\end{equation}
\begin{equation}
    p_t^{eff}(r)=\frac{2\mathcal{H}_1r^6+\kappa^{\alpha-1}r^{6\alpha}(2\mathcal{G}_2+\mathcal{G}_3)}{2\kappa^\alpha r^{6\alpha+6}r_0},
\end{equation}
where we have defined the parameters $\mathcal{G}_1=3r_0^2r^2-2r_0^3r+(r-2r_0)(\mathcal{E}_1-r_0^2r)$, $ \mathcal{G}_2=(r_0r^2-\mathcal{E}_1)(r+r_0)r_0$ and $\mathcal{G}_3=\mathcal{E}_2(r-r_0)r$.
\subsubsection{\texorpdfstring{$\text{exp}(2\Phi(r))=1+\gamma^2/r^2$}{}}
Here $\gamma$ is some parameter. The pressure components are:
\begin{equation}
    p_r^{eff}(r)=\frac{\mathcal{H}_1r^4(r^2+\gamma^2)-\kappa^{\alpha-1}r^{6\alpha}\mathcal{J}_1}{\kappa^\alpha r^{6\alpha+4}(r^2+\gamma^2)r_0},
\end{equation}
\begin{equation}
    p_t^{eff}(r)=\frac{2\mathcal{H}_1r^4{(r^2+\gamma^2)}^2+\kappa^{\alpha-1}r^{6\alpha}(2\mathcal{J}_2+\mathcal{J}_3)}{2\kappa^\alpha r^{6\alpha+4}{(r^2+\gamma^2)}^2r_0},
\end{equation}
where $ \mathcal{J}_1=2\gamma^2r_0r^2+(r^2-\gamma^2)\mathcal{E}_1$, $ \mathcal{J}_2=\gamma^2(\gamma^2+2r^2)(r_0r^2-\mathcal{E}_1)$, and $\mathcal{J}_3=(\gamma^2+r^2)r^2\mathcal{E}_2$.\\
With the results established above, we can now inspect the outcome of the energy conditions, visualize and verify its compliance, specifically the NEC. Classically, the NEC is defined as $\mathcal{T}^{eff}_{\mu\nu} k^\mu k^\nu \geq 0$, i.e., 
\begin{equation}
\rho^{eff}+p_r^{eff}\geq 0,
\end{equation}
where $k^\mu$ represents a null vector.
We observe from Fig.\ref{fig:NECrM1} that the NEC is disobeyed, as $\rho^{eff}+p_r^{eff}$ is negative at the throat $r_0=1$ for all cases. This implies violation of all classical energy conditions, as NEC is the weakest of them all.
\begin{figure*}
    \includegraphics[scale=0.5]{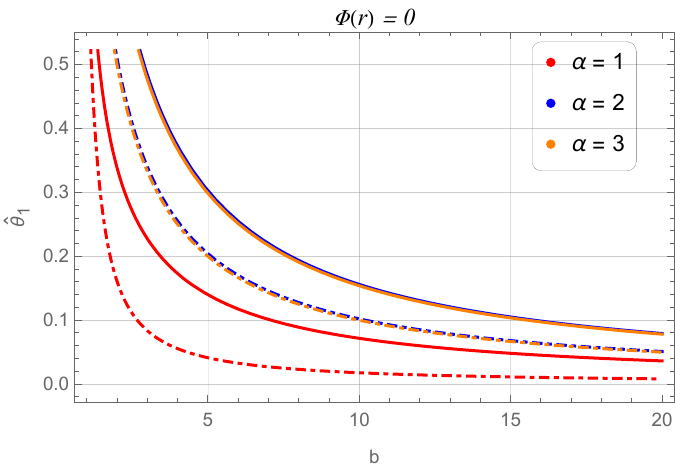}
    \includegraphics[scale=0.5]{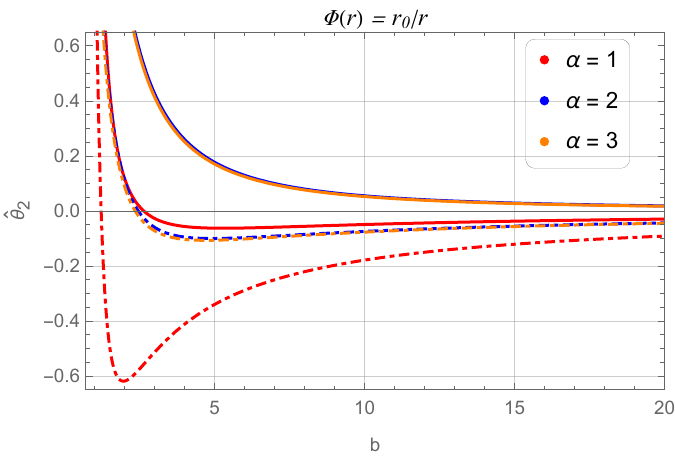}
   \includegraphics[scale=0.5]{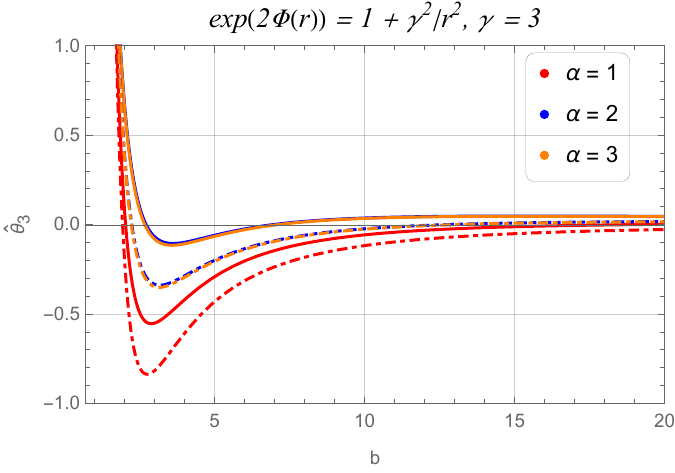}
    \caption{Plots for deflection angles against the impact parameter. Dashed curves are for the uncharged case, while solid curves correspond to $Q=0.765$. We have chosen the KMM case for representation, with $\lambda=0.1$ and $r_0=1$.}
    \label{fig:defl angle} 
\end{figure*}
\section{Gravitational Lensing Effect by charged GUP Casimir Wormholes}\label{sec: Lensing}
In this section, we shall use the Gauss-Bonnet Theorem (GBT) approach to calculate the deflection angle due to gravitational lensing. Let $\mathcal{A}_R$ be a non-singular, connected domain bounded by circular curve $\gamma_R$ and the geodesic $\gamma_{\bar{g}}$ i.e. $\partial\mathcal{A}_R=\gamma_R\cup\gamma_{\bar{g}}$, with the optical metric $\bar{g}$. Suppose $K$ denotes the Gaussian curvature and $\kappa_g$ is the geodesic curvature. The GBT provides a fundamental relation between geometry and the topology of a manifold, by stating the following \cite{gibbons2008applications}:
\begin{equation}
    \iint_{\mathcal{A}_R}KdS+\oint_{\partial\mathcal{A}_R}\kappa_gdt+\sum_{i}\theta_i=2\pi\chi(\mathcal{A}_R).
\end{equation}
Here, $dS$ is the optical surface element, $\theta_i$ denotes the exterior angle at the $i^{th}$ vertex, while $\chi(\mathcal{A}_R)$ is the Euler characteristic number for the surface $\mathcal{A}_R$. By definition, $\kappa_g(\gamma_{\bar{g}})=0$, thus the contribution from the domain's boundary is only from $\gamma_R$ whose geodesic curvature is given by
\begin{equation}
    \kappa_g(\gamma_R)=|\nabla_{\dot{\gamma}_R}\dot{\gamma}_R|.
\end{equation}
Note that we defined $\gamma_R:=r(\phi)=R=\text{constant}$. By using the unit speed condition, one may find that $\kappa_g(\gamma_R)\to1/R$ for $R\to\infty$. As a consequence, we get the following result
\begin{equation}
    \lim_{R\to\infty}\kappa_g(\gamma_R)\frac{dt}{d\phi}=1.
\end{equation}
Furthermore, at large distances, the sum of the jump angles at the source $S$ and observer $O$ is $\theta_S+\theta_O\to\pi$. For our construction, the GBT yields
\begin{equation}
    \lim_{R\to\infty}\Biggl[\iint_{\mathcal{A}_R}KdS+\int_{0}^{\pi+\bm{\hat{\theta}}}{\left(\kappa_g\frac{dt}{d\phi}\right)}_{\gamma_R}d\phi\Biggr]=\pi.
\end{equation}
Thus, the deflection angle of a light ray can be calculated as
\begin{equation}
    \bm{\hat{\theta}}=-\int_0^\pi\int_{r=\frac{\text{b}}{\text{sin}\phi}}^\infty K\sqrt{\text{det}(\bar{g})}drd\phi,
\end{equation}
where `$\text{b}$' is the impact parameter.\\
Using the method to calculate the Gaussian curvature for a given spacetime metric, detailed in \cite{gibbons2008applications}, for the metric \eqref{metric}
\begin{equation}
    K=e^{2\Phi}\left(1-\frac{b}{r}\right)\Biggl[\Phi''+\frac{\Phi'}{r}-\frac{rb'-b}{2r(r-b)}\left(\Phi'-\frac{1}{r}\right)\Biggr].
\end{equation}
We now evaluate the deflection angles for each case of gravitational redshift. Unlike in the past works we shall include the contributions from the GUP terms, as they are the crux of our study. This section deals with the epilogue of the works \cite{jusufi2020traversable,samart2022charged}.\\
We first begin with the simplest case, a spacetime with the absence of tidal forces, i.e., $\Phi(r)=0$. Hence, the deflection of light rays is due to spatial lensing more so than gravitational tides. The deflection angle is found to be
\begin{multline}
    \bm{\hat{\theta}_1}\simeq\frac{R_1}{\imp}+\frac{r_2^2}{r_0\imp}\left(1-\frac{\pi r_0}{4\imp}\right)-\frac{\sqrt{\pi}\Gamma(2\alpha-\sfrac{3}{2}){(-r_1^2)}^\alpha}{4\kappa^{\alpha-1}\Gamma(2\alpha-1)\imp^{4\alpha-2}}\times\\
    \Biggl[1+\frac{(4\alpha-3)r_i^2}{4\imp^2}\Biggr],
\end{multline}
by defining the term $R_1$ as
\begin{equation}
    R_1=r_0-\frac{2\alpha-1}{\kappa^{\alpha-1}}{(-r_1^2)}^\alpha\Biggl[\frac{r_0^{3-4\alpha}}{3-4\alpha}+\frac{\alpha r_i^2r_0^{1-4\alpha}}{1-4\alpha}\Biggr].
\end{equation}
Next for the redshift $\Phi(r)=r_0/r$, the deflection angle can be read as
\begin{multline}
    \bm{\hat{\theta}_2}\simeq-\frac{2r_0}{\imp}+\frac{R_1}{\imp}\left(1+\frac{3\pi r_0}{4\imp}\right)+\frac{2r_2^2}{r_0\imp}\left(1+\frac{3\pi r_0}{8\imp}-\frac{4r_0^2}{3\imp^2}\right)\\
    -\frac{2\alpha-1}{\kappa^{\alpha-1}}{(-r_1^2)}^\alpha\Biggl\{\frac{\sqrt{\pi}\Gamma(2\alpha-\sfrac{3}{2})}{4\Gamma(2\alpha)\imp^{4\alpha-2}}\Biggl[1+\frac{8(3\alpha-1)\Gamma(2\alpha)^2}{(4\alpha-1)^2\Gamma(2\alpha-\sfrac{1}{2})^2}\\\frac{r_0}{\imp}+\frac{4\alpha-1}{2\alpha}\frac{r_0^2}{\imp^2}\Biggr]
    +\frac{\alpha\sqrt{\pi}r_i^2\Gamma(2\alpha-\sfrac{1}{2})}{4\Gamma(2\alpha+1)\imp^{4\alpha}}\times\\
    \Biggl[1+\frac{4(6\alpha+1)\Gamma(2\alpha+1)^2}{(4\alpha+1)^2\Gamma(2\alpha+\sfrac{1}{2})^2}\frac{r_0}{\imp}+\frac{4\alpha+1}{2\alpha+1}\frac{r_0^2}{\imp^2}\Biggr]\Biggr\}.
\end{multline}
Finally, for the $\text{exp}(2\Phi(r))=1+\gamma^2/r^2$ case, we have obtained the following expression for the deflection angle
\begin{multline}\label{def3}
    \bm{\hat{\theta}_3}\simeq\bm{\hat{\theta}_1}-\gamma^2\Biggl\{\frac{\pi}{2\imp^2}-\frac{4R_1}{3\imp^3}-\frac{3\pi\gamma^2}{16\imp^4}+\frac{32\gamma^2R_1}{75\imp^5}\\
    -r_2^2\left(\frac{4}{3r_0\imp^3}-\frac{3\pi}{8\imp^4}-\frac{32\gamma^2}{75r_0\imp^5}+\frac{5\pi\gamma^2}{48\imp^6}\right)+\frac{2\alpha-1}{\kappa^{\alpha-1}}{(-r_1^2)}^\alpha\times\\
    \Biggl[\frac{\sqrt{\pi}\Gamma(2\alpha+\sfrac{1}{2})}{(4\alpha-3)\Gamma(2\alpha+1)b^{4\alpha}}\left(1-\frac{\gamma^2(4\alpha+1)}{2(2\alpha+1)^2\imp^2}\right)\\
    +\frac{\alpha\sqrt{\pi}\Gamma(2\alpha+\sfrac{3}{2})r_i^2}{(4\alpha-1)\Gamma(2\alpha+2)\imp^{4\alpha+2}}\left(1-\frac{\gamma^2(4\alpha+3)}{8(\alpha+1)^2\imp^2}\right)\Biggr]\Biggr\}.
\end{multline}
Note that the above case \eqref{def3} reduces to the result for zero redshift when $\gamma$ approaches zero.
The deflection angle for each case is illustrated as a function of the impact parameter `$\imp$' in Fig.\ref{fig:defl angle}. We notice that the parameter $\gamma$ in the third redshift model plays a significant role in lensing. Fig.\ref{fig:defl angle} also demonstrates that the deflection angle approaches zero as the impact parameter extends towards infinity. This implies that, the light rays moving far away from the wormhole, where the gravitational influence of the wormhole is minimal, remains almost unaffected from its trajectory. Conversely, when `$\imp$' is close to the wormhole throat, the deflection angle increases significantly and is positive. This means that the light rays sufficiently close to the source of the gravitational potential are more likely to fall toward the wormhole.\\
\section{Conclusions}\label{sec: conclusions}
In this letter, we have studied the GUP-corrected Casimir wormholes in the Curvature-Lagrangian coupled $f(R,\Lagr)$ framework by solving the modified counterpart of the Einstein-Maxwell field equations. In particular, we have considered $f(R,\Lagr)=\frac{R}{2\kappa}+\Lagr^{\,\,\alpha}$ background. Although an exact, well-defined, analytic solution exists for the wormhole geometry, its corresponding redshift function does not have an analytic solution; therefore we have implemented three widely used redshift function models, namely $\Phi(r)=0$, $\Phi(r)=\frac{r_0}{r}$ and $\text{exp}(2\Phi(r))=1+\frac{\gamma^2}{r^2}$, and calculated the components of pressure experienced by a test object. It was observed that the asymptotic flatness condition for the shape function holds only if $\alpha>1/2$, in addition, $b(r)$ is real-valued only if $\alpha$ is an integer. As a consequence of the flare-out condition, we obtain a bound on the allowable electric charge in natural units to be $|Q|<0.7656$ for the KMM case. Clearly, from Fig.\ref{fig:embedding}, modifying GR and the presence of an electric charge affects the wormhole geometry by elongating the throat.
Furthermore, we have checked the NEC for each redshift model, which was found to be violated at $r=r_0$, indicating the presence of exotic matter at the throat, an inseparable requirement for the traversability of the wormhole. Thus, we can infer that all the classical energy conditions are also violated.\\
\indent Lastly, we employ the GBT to compute the deflection angles for each redshift function under consideration. It is important to note that the redshift function significantly impacts the deflection of light rays. We have included higher-order terms, GUP effects, and the effect of electric charge in order to arrive at a more nuanced conclusion. We noticed that light rays bend towards the wormhole if there are no tidal forces. However, the results vary for the other two redshift functions; light may diverge for certain impact parameter values, and in some cases, it remains unaffected. Regardless of the kind of potential in action, it is found that when the impact parameter is very close to the wormhole throat, the deviation of trajectory is steeper and towards the wormhole. Although the model considered in this letter is compliant with cosmological observations \cite{kavya2022constraining}, one may further explore this subject by utilizing a model with non-minimal couplings, such as in \cite{solanki2023wormhole}, or the exponential model in \cite{harko2010f} to acquire a more generalized perspective.\\
\textbf{Data availability:} There are no new data associated with this article.

\acknowledgments
ZH acknowledges the Department of Science and Technology (DST), Government of India, New Delhi, for awarding a Senior Research Fellowship (File No. DST/INSPIRE Fellowship/2019/IF190911). PKS acknowledges the National Board for Higher Mathematics (NBHM) under the Department of Atomic Energy (DAE), Govt. of India, for financial support to carry out the Research project No.: 02011/3/2022 NBHM(R.P.)/R\&D II/2152 Dt.14.02.2022. 
\bibliography{ref.bib}

\begin{thebibliography}{46}%
\makeatletter
\providecommand \@ifxundefined [1]{%
 \@ifx{#1\undefined}
}%
\providecommand \@ifnum [1]{%
 \ifnum #1\expandafter \@firstoftwo
 \else \expandafter \@secondoftwo
 \fi
}%
\providecommand \@ifx [1]{%
 \ifx #1\expandafter \@firstoftwo
 \else \expandafter \@secondoftwo
 \fi
}%
\providecommand \natexlab [1]{#1}%
\providecommand \enquote  [1]{``#1''}%
\providecommand \bibnamefont  [1]{#1}%
\providecommand \bibfnamefont [1]{#1}%
\providecommand \citenamefont [1]{#1}%
\providecommand \href@noop [0]{\@secondoftwo}%
\providecommand \href [0]{\begingroup \@sanitize@url \@href}%
\providecommand \@href[1]{\@@startlink{#1}\@@href}%
\providecommand \@@href[1]{\endgroup#1\@@endlink}%
\providecommand \@sanitize@url [0]{\catcode `\\12\catcode `\$12\catcode `\&12\catcode `\#12\catcode `\^12\catcode `\_12\catcode `\%12\relax}%
\providecommand \@@startlink[1]{}%
\providecommand \@@endlink[0]{}%
\providecommand \url  [0]{\begingroup\@sanitize@url \@url }%
\providecommand \@url [1]{\endgroup\@href {#1}{\urlprefix }}%
\providecommand \urlprefix  [0]{URL }%
\providecommand \Eprint [0]{\href }%
\providecommand \doibase [0]{https://doi.org/}%
\providecommand \selectlanguage [0]{\@gobble}%
\providecommand \bibinfo  [0]{\@secondoftwo}%
\providecommand \bibfield  [0]{\@secondoftwo}%
\providecommand \translation [1]{[#1]}%
\providecommand \BibitemOpen [0]{}%
\providecommand \bibitemStop [0]{}%
\providecommand \bibitemNoStop [0]{.\EOS\space}%
\providecommand \EOS [0]{\spacefactor3000\relax}%
\providecommand \BibitemShut  [1]{\csname bibitem#1\endcsname}%
\let\auto@bib@innerbib\@empty
\bibitem [{\citenamefont {Flamm}(1916)}]{flamm1916beitrage}%
  \BibitemOpen
  \bibfield  {author} {\bibinfo {author} {\bibfnamefont {L.}~\bibnamefont {Flamm}},\ }\href@noop {} {\emph {\bibinfo {title} {Beitr{\"a}ge zur Einsteinschen gravitationstheorie}}}\ (\bibinfo  {publisher} {Hirzel},\ \bibinfo {year} {1916})\BibitemShut {NoStop}%
\bibitem [{\citenamefont {Einstein}\ and\ \citenamefont {Rosen}(1935)}]{einstein1935particle}%
  \BibitemOpen
  \bibfield  {author} {\bibinfo {author} {\bibfnamefont {A.}~\bibnamefont {Einstein}}\ and\ \bibinfo {author} {\bibfnamefont {N.}~\bibnamefont {Rosen}},\ }\href@noop {} {\bibfield  {journal} {\bibinfo  {journal} {Physical Review}\ }\textbf {\bibinfo {volume} {48}},\ \bibinfo {pages} {73} (\bibinfo {year} {1935})}\BibitemShut {NoStop}%
\bibitem [{\citenamefont {Morris}\ and\ \citenamefont {Thorne}(1988)}]{morris1988wormholes}%
  \BibitemOpen
  \bibfield  {author} {\bibinfo {author} {\bibfnamefont {M.~S.}\ \bibnamefont {Morris}}\ and\ \bibinfo {author} {\bibfnamefont {K.~S.}\ \bibnamefont {Thorne}},\ }\href@noop {} {\bibfield  {journal} {\bibinfo  {journal} {American Journal of Physics}\ }\textbf {\bibinfo {volume} {56}},\ \bibinfo {pages} {395} (\bibinfo {year} {1988})}\BibitemShut {NoStop}%
\bibitem [{\citenamefont {Casimir}(1948)}]{casimir1948attraction}%
  \BibitemOpen
  \bibfield  {author} {\bibinfo {author} {\bibfnamefont {H.~B.}\ \bibnamefont {Casimir}},\ }in\ \href@noop {} {\emph {\bibinfo {booktitle} {Proc. Kon. Ned. Akad. Wet.}}},\ Vol.~\bibinfo {volume} {51}\ (\bibinfo {year} {1948})\ p.\ \bibinfo {pages} {793}\BibitemShut {NoStop}%
\bibitem [{\citenamefont {Dzyaloshinskii}\ \emph {et~al.}(1961)\citenamefont {Dzyaloshinskii}, \citenamefont {Lifshitz},\ and\ \citenamefont {Pitaevskii}}]{dzyaloshinskii1961general}%
  \BibitemOpen
  \bibfield  {author} {\bibinfo {author} {\bibfnamefont {I.~E.}\ \bibnamefont {Dzyaloshinskii}}, \bibinfo {author} {\bibfnamefont {E.~M.}\ \bibnamefont {Lifshitz}},\ and\ \bibinfo {author} {\bibfnamefont {L.~P.}\ \bibnamefont {Pitaevskii}},\ }\href@noop {} {\bibfield  {journal} {\bibinfo  {journal} {Advances in Physics}\ }\textbf {\bibinfo {volume} {10}},\ \bibinfo {pages} {165} (\bibinfo {year} {1961})}\BibitemShut {NoStop}%
\bibitem [{\citenamefont {Black}\ \emph {et~al.}(1960)\citenamefont {Black}, \citenamefont {De~Jongh}, \citenamefont {Overbeek},\ and\ \citenamefont {Sparnaay}}]{black1960measurements}%
  \BibitemOpen
  \bibfield  {author} {\bibinfo {author} {\bibfnamefont {W.}~\bibnamefont {Black}}, \bibinfo {author} {\bibfnamefont {J.}~\bibnamefont {De~Jongh}}, \bibinfo {author} {\bibfnamefont {J.~T.~G.}\ \bibnamefont {Overbeek}},\ and\ \bibinfo {author} {\bibfnamefont {M.}~\bibnamefont {Sparnaay}},\ }\href@noop {} {\bibfield  {journal} {\bibinfo  {journal} {Transactions of the Faraday Society}\ }\textbf {\bibinfo {volume} {56}},\ \bibinfo {pages} {1597} (\bibinfo {year} {1960})}\BibitemShut {NoStop}%
\bibitem [{\citenamefont {Lamoreaux}(1997)}]{lamoreaux1997demonstration}%
  \BibitemOpen
  \bibfield  {author} {\bibinfo {author} {\bibfnamefont {S.~K.}\ \bibnamefont {Lamoreaux}},\ }\href@noop {} {\bibfield  {journal} {\bibinfo  {journal} {Physical Review Letters}\ }\textbf {\bibinfo {volume} {78}},\ \bibinfo {pages} {5} (\bibinfo {year} {1997})}\BibitemShut {NoStop}%
\bibitem [{\citenamefont {Bressi}\ \emph {et~al.}(2002)\citenamefont {Bressi}, \citenamefont {Carugno}, \citenamefont {Onofrio},\ and\ \citenamefont {Ruoso}}]{bressi2002measurement}%
  \BibitemOpen
  \bibfield  {author} {\bibinfo {author} {\bibfnamefont {G.}~\bibnamefont {Bressi}}, \bibinfo {author} {\bibfnamefont {G.}~\bibnamefont {Carugno}}, \bibinfo {author} {\bibfnamefont {R.}~\bibnamefont {Onofrio}},\ and\ \bibinfo {author} {\bibfnamefont {G.}~\bibnamefont {Ruoso}},\ }\href@noop {} {\bibfield  {journal} {\bibinfo  {journal} {Physical review letters}\ }\textbf {\bibinfo {volume} {88}},\ \bibinfo {pages} {041804} (\bibinfo {year} {2002})}\BibitemShut {NoStop}%
\bibitem [{\citenamefont {Kuhfittig}(2011)}]{kuhfittig2011feasibility}%
  \BibitemOpen
  \bibfield  {author} {\bibinfo {author} {\bibfnamefont {P.~K.}\ \bibnamefont {Kuhfittig}},\ }\href@noop {} {\bibfield  {journal} {\bibinfo  {journal} {Central European Journal of Physics}\ }\textbf {\bibinfo {volume} {9}},\ \bibinfo {pages} {1144} (\bibinfo {year} {2011})}\BibitemShut {NoStop}%
\bibitem [{\citenamefont {Kim}\ and\ \citenamefont {Lee}(2001)}]{kim2001exact}%
  \BibitemOpen
  \bibfield  {author} {\bibinfo {author} {\bibfnamefont {S.-W.}\ \bibnamefont {Kim}}\ and\ \bibinfo {author} {\bibfnamefont {H.}~\bibnamefont {Lee}},\ }\href@noop {} {\bibfield  {journal} {\bibinfo  {journal} {Physical Review D}\ }\textbf {\bibinfo {volume} {63}},\ \bibinfo {pages} {064014} (\bibinfo {year} {2001})}\BibitemShut {NoStop}%
\bibitem [{\citenamefont {Capozziello}\ \emph {et~al.}(2021)\citenamefont {Capozziello}, \citenamefont {Luongo},\ and\ \citenamefont {Mauro}}]{capozziello2021traversable}%
  \BibitemOpen
  \bibfield  {author} {\bibinfo {author} {\bibfnamefont {S.}~\bibnamefont {Capozziello}}, \bibinfo {author} {\bibfnamefont {O.}~\bibnamefont {Luongo}},\ and\ \bibinfo {author} {\bibfnamefont {L.}~\bibnamefont {Mauro}},\ }\href@noop {} {\bibfield  {journal} {\bibinfo  {journal} {The European Physical Journal Plus}\ }\textbf {\bibinfo {volume} {136}},\ \bibinfo {pages} {1} (\bibinfo {year} {2021})}\BibitemShut {NoStop}%
\bibitem [{\citenamefont {De~Falco}\ \emph {et~al.}(2021{\natexlab{a}})\citenamefont {De~Falco}, \citenamefont {Battista}, \citenamefont {Capozziello},\ and\ \citenamefont {De~Laurentis}}]{de2021reconstructing}%
  \BibitemOpen
  \bibfield  {author} {\bibinfo {author} {\bibfnamefont {V.}~\bibnamefont {De~Falco}}, \bibinfo {author} {\bibfnamefont {E.}~\bibnamefont {Battista}}, \bibinfo {author} {\bibfnamefont {S.}~\bibnamefont {Capozziello}},\ and\ \bibinfo {author} {\bibfnamefont {M.}~\bibnamefont {De~Laurentis}},\ }\href@noop {} {\bibfield  {journal} {\bibinfo  {journal} {The European Physical Journal C}\ }\textbf {\bibinfo {volume} {81}},\ \bibinfo {pages} {1} (\bibinfo {year} {2021}{\natexlab{a}})}\BibitemShut {NoStop}%
\bibitem [{\citenamefont {Capozziello}\ and\ \citenamefont {Godani}(2022)}]{Capozziello:2022zoz}%
  \BibitemOpen
  \bibfield  {author} {\bibinfo {author} {\bibfnamefont {S.}~\bibnamefont {Capozziello}}\ and\ \bibinfo {author} {\bibfnamefont {N.}~\bibnamefont {Godani}},\ }\href@noop {} {\bibfield  {journal} {\bibinfo  {journal} {Phys. Lett. B}\ }\textbf {\bibinfo {volume} {835}},\ \bibinfo {pages} {137572} (\bibinfo {year} {2022})}\BibitemShut {NoStop}%
\bibitem [{\citenamefont {De~Falco}\ \emph {et~al.}(2021{\natexlab{b}})\citenamefont {De~Falco}, \citenamefont {Battista}, \citenamefont {Capozziello},\ and\ \citenamefont {De~Laurentis}}]{de2021testing}%
  \BibitemOpen
  \bibfield  {author} {\bibinfo {author} {\bibfnamefont {V.}~\bibnamefont {De~Falco}}, \bibinfo {author} {\bibfnamefont {E.}~\bibnamefont {Battista}}, \bibinfo {author} {\bibfnamefont {S.}~\bibnamefont {Capozziello}},\ and\ \bibinfo {author} {\bibfnamefont {M.}~\bibnamefont {De~Laurentis}},\ }\href@noop {} {\bibfield  {journal} {\bibinfo  {journal} {Physical Review D}\ }\textbf {\bibinfo {volume} {103}},\ \bibinfo {pages} {044007} (\bibinfo {year} {2021}{\natexlab{b}})}\BibitemShut {NoStop}%
\bibitem [{\citenamefont {Garattini}(2019)}]{garattini2019casimir}%
  \BibitemOpen
  \bibfield  {author} {\bibinfo {author} {\bibfnamefont {R.}~\bibnamefont {Garattini}},\ }\href@noop {} {\bibfield  {journal} {\bibinfo  {journal} {The European Physical Journal C}\ }\textbf {\bibinfo {volume} {79}},\ \bibinfo {pages} {951} (\bibinfo {year} {2019})}\BibitemShut {NoStop}%
\bibitem [{\citenamefont {Garattini}(2023)}]{garattini2023effects}%
  \BibitemOpen
  \bibfield  {author} {\bibinfo {author} {\bibfnamefont {R.}~\bibnamefont {Garattini}},\ }\href@noop {} {\bibfield  {journal} {\bibinfo  {journal} {The European Physical Journal C}\ }\textbf {\bibinfo {volume} {83}},\ \bibinfo {pages} {369} (\bibinfo {year} {2023})}\BibitemShut {NoStop}%
\bibitem [{\citenamefont {Jusufi}\ \emph {et~al.}(2020)\citenamefont {Jusufi}, \citenamefont {Channuie},\ and\ \citenamefont {Jamil}}]{jusufi2020traversable}%
  \BibitemOpen
  \bibfield  {author} {\bibinfo {author} {\bibfnamefont {K.}~\bibnamefont {Jusufi}}, \bibinfo {author} {\bibfnamefont {P.}~\bibnamefont {Channuie}},\ and\ \bibinfo {author} {\bibfnamefont {M.}~\bibnamefont {Jamil}},\ }\href@noop {} {\bibfield  {journal} {\bibinfo  {journal} {The European Physical Journal C}\ }\textbf {\bibinfo {volume} {80}},\ \bibinfo {pages} {127} (\bibinfo {year} {2020})}\BibitemShut {NoStop}%
\bibitem [{\citenamefont {Samart}\ \emph {et~al.}(2022)\citenamefont {Samart}, \citenamefont {Tangphati},\ and\ \citenamefont {Channuie}}]{samart2022charged}%
  \BibitemOpen
  \bibfield  {author} {\bibinfo {author} {\bibfnamefont {D.}~\bibnamefont {Samart}}, \bibinfo {author} {\bibfnamefont {T.}~\bibnamefont {Tangphati}},\ and\ \bibinfo {author} {\bibfnamefont {P.}~\bibnamefont {Channuie}},\ }\href@noop {} {\bibfield  {journal} {\bibinfo  {journal} {Nuclear Physics B}\ }\textbf {\bibinfo {volume} {980}},\ \bibinfo {pages} {115848} (\bibinfo {year} {2022})}\BibitemShut {NoStop}%
\bibitem [{\citenamefont {Battista}\ \emph {et~al.}(2024)\citenamefont {Battista}, \citenamefont {Capozziello},\ and\ \citenamefont {Errehymy}}]{battista2024generalized}%
  \BibitemOpen
  \bibfield  {author} {\bibinfo {author} {\bibfnamefont {E.}~\bibnamefont {Battista}}, \bibinfo {author} {\bibfnamefont {S.}~\bibnamefont {Capozziello}},\ and\ \bibinfo {author} {\bibfnamefont {A.}~\bibnamefont {Errehymy}},\ }\href@noop {} {\bibfield  {journal} {\bibinfo  {journal} {arXiv preprint arXiv:2409.09750}\ } (\bibinfo {year} {2024})}\BibitemShut {NoStop}%
\bibitem [{\citenamefont {Hassan}\ \emph {et~al.}(2022)\citenamefont {Hassan}, \citenamefont {Ghosh}, \citenamefont {Sahoo},\ and\ \citenamefont {Bamba}}]{ZinnatGUP1}%
  \BibitemOpen
  \bibfield  {author} {\bibinfo {author} {\bibfnamefont {Z.}~\bibnamefont {Hassan}}, \bibinfo {author} {\bibfnamefont {S.}~\bibnamefont {Ghosh}}, \bibinfo {author} {\bibfnamefont {P.~K.}\ \bibnamefont {Sahoo}},\ and\ \bibinfo {author} {\bibfnamefont {K.}~\bibnamefont {Bamba}},\ }\href@noop {} {\bibfield  {journal} {\bibinfo  {journal} {The European Physical Journal C}\ }\textbf {\bibinfo {volume} {82}},\ \bibinfo {pages} {1116} (\bibinfo {year} {2022})}\BibitemShut {NoStop}%
\bibitem [{\citenamefont {Hassan}\ \emph {et~al.}(2023)\citenamefont {Hassan}, \citenamefont {Ghosh}, \citenamefont {Sahoo},\ and\ \citenamefont {Rao}}]{ZinnatGUP2}%
  \BibitemOpen
  \bibfield  {author} {\bibinfo {author} {\bibfnamefont {Z.}~\bibnamefont {Hassan}}, \bibinfo {author} {\bibfnamefont {S.}~\bibnamefont {Ghosh}}, \bibinfo {author} {\bibfnamefont {P.~K.}\ \bibnamefont {Sahoo}},\ and\ \bibinfo {author} {\bibfnamefont {V.~S.~H.}\ \bibnamefont {Rao}},\ }\href@noop {} {\bibfield  {journal} {\bibinfo  {journal} {General Relativity and Gravitation}\ }\textbf {\bibinfo {volume} {55}},\ \bibinfo {pages} {90} (\bibinfo {year} {2023})}\BibitemShut {NoStop}%
\bibitem [{\citenamefont {Khatri}\ and\ \citenamefont {Lalvohbika}(2024)}]{khatri2024casimir}%
  \BibitemOpen
  \bibfield  {author} {\bibinfo {author} {\bibfnamefont {M.}~\bibnamefont {Khatri}}\ and\ \bibinfo {author} {\bibfnamefont {J.}~\bibnamefont {Lalvohbika}},\ }\href@noop {} {\bibfield  {journal} {\bibinfo  {journal} {Chinese Journal of Physics}\ }\textbf {\bibinfo {volume} {89}},\ \bibinfo {pages} {1222} (\bibinfo {year} {2024})}\BibitemShut {NoStop}%
\bibitem [{\citenamefont {Harko}\ and\ \citenamefont {Lobo}(2010)}]{harko2010f}%
  \BibitemOpen
  \bibfield  {author} {\bibinfo {author} {\bibfnamefont {T.}~\bibnamefont {Harko}}\ and\ \bibinfo {author} {\bibfnamefont {F.~S.}\ \bibnamefont {Lobo}},\ }\href@noop {} {\bibfield  {journal} {\bibinfo  {journal} {The European Physical Journal C}\ }\textbf {\bibinfo {volume} {70}},\ \bibinfo {pages} {373} (\bibinfo {year} {2010})}\BibitemShut {NoStop}%
\bibitem [{\citenamefont {Faraoni}(2004)}]{faraoni2004scalar}%
  \BibitemOpen
  \bibfield  {author} {\bibinfo {author} {\bibfnamefont {V.}~\bibnamefont {Faraoni}},\ }\href@noop {} {\emph {\bibinfo {title} {Cosmology in Scalar-Tensor Gravity}}}\ (\bibinfo  {publisher} {Springer},\ \bibinfo {year} {2004})\BibitemShut {NoStop}%
\bibitem [{\citenamefont {Bertolami}\ \emph {et~al.}(2008)\citenamefont {Bertolami}, \citenamefont {P{\'a}ramos},\ and\ \citenamefont {Turyshev}}]{bertolami2008general}%
  \BibitemOpen
  \bibfield  {author} {\bibinfo {author} {\bibfnamefont {O.}~\bibnamefont {Bertolami}}, \bibinfo {author} {\bibfnamefont {J.}~\bibnamefont {P{\'a}ramos}},\ and\ \bibinfo {author} {\bibfnamefont {S.~G.}\ \bibnamefont {Turyshev}},\ }in\ \href@noop {} {\emph {\bibinfo {booktitle} {Lasers, Clocks and Drag-Free Control: Exploration of Relativistic Gravity in Space}}}\ (\bibinfo  {publisher} {Springer},\ \bibinfo {year} {2008})\ pp.\ \bibinfo {pages} {27--74}\BibitemShut {NoStop}%
\bibitem [{\citenamefont {Lobato}\ \emph {et~al.}(2021)\citenamefont {Lobato}, \citenamefont {Carvalho},\ and\ \citenamefont {Bertulani}}]{lobato2021neutron}%
  \BibitemOpen
  \bibfield  {author} {\bibinfo {author} {\bibfnamefont {R.}~\bibnamefont {Lobato}}, \bibinfo {author} {\bibfnamefont {G.}~\bibnamefont {Carvalho}},\ and\ \bibinfo {author} {\bibfnamefont {C.}~\bibnamefont {Bertulani}},\ }\href@noop {} {\bibfield  {journal} {\bibinfo  {journal} {The European Physical Journal C}\ }\textbf {\bibinfo {volume} {81}},\ \bibinfo {pages} {1} (\bibinfo {year} {2021})}\BibitemShut {NoStop}%
\bibitem [{\citenamefont {Harko}\ and\ \citenamefont {Shahidi}(2022)}]{harko2022palatini}%
  \BibitemOpen
  \bibfield  {author} {\bibinfo {author} {\bibfnamefont {T.}~\bibnamefont {Harko}}\ and\ \bibinfo {author} {\bibfnamefont {S.}~\bibnamefont {Shahidi}},\ }\href@noop {} {\bibfield  {journal} {\bibinfo  {journal} {The European Physical Journal C}\ }\textbf {\bibinfo {volume} {82}},\ \bibinfo {pages} {1003} (\bibinfo {year} {2022})}\BibitemShut {NoStop}%
\bibitem [{\citenamefont {Harko}\ and\ \citenamefont {Lake}(2015)}]{harko2015cosmic}%
  \BibitemOpen
  \bibfield  {author} {\bibinfo {author} {\bibfnamefont {T.}~\bibnamefont {Harko}}\ and\ \bibinfo {author} {\bibfnamefont {M.~J.}\ \bibnamefont {Lake}},\ }\href@noop {} {\bibfield  {journal} {\bibinfo  {journal} {The European Physical Journal C}\ }\textbf {\bibinfo {volume} {75}},\ \bibinfo {pages} {1} (\bibinfo {year} {2015})}\BibitemShut {NoStop}%
\bibitem [{\citenamefont {Solanki}\ \emph {et~al.}(2023)\citenamefont {Solanki}, \citenamefont {Hassan},\ and\ \citenamefont {Sahoo}}]{solanki2023wormhole}%
  \BibitemOpen
  \bibfield  {author} {\bibinfo {author} {\bibfnamefont {R.}~\bibnamefont {Solanki}}, \bibinfo {author} {\bibfnamefont {Z.}~\bibnamefont {Hassan}},\ and\ \bibinfo {author} {\bibfnamefont {P.~K.}\ \bibnamefont {Sahoo}},\ }\href@noop {} {\bibfield  {journal} {\bibinfo  {journal} {Chinese Journal of Physics}\ }\textbf {\bibinfo {volume} {85}},\ \bibinfo {pages} {74} (\bibinfo {year} {2023})}\BibitemShut {NoStop}%
\bibitem [{\citenamefont {Kavya}\ \emph {et~al.}(2023)\citenamefont {Kavya}, \citenamefont {Venkatesha}, \citenamefont {Mustafa},\ and\ \citenamefont {Sahoo}}]{kavya2023possible}%
  \BibitemOpen
  \bibfield  {author} {\bibinfo {author} {\bibfnamefont {N.}~\bibnamefont {Kavya}}, \bibinfo {author} {\bibfnamefont {V.}~\bibnamefont {Venkatesha}}, \bibinfo {author} {\bibfnamefont {G.}~\bibnamefont {Mustafa}},\ and\ \bibinfo {author} {\bibfnamefont {P.~K.}\ \bibnamefont {Sahoo}},\ }\href@noop {} {\bibfield  {journal} {\bibinfo  {journal} {Annals of Physics}\ }\textbf {\bibinfo {volume} {455}},\ \bibinfo {pages} {169383} (\bibinfo {year} {2023})}\BibitemShut {NoStop}%
\bibitem [{\citenamefont {Ta{\d s}er}\ \emph {et~al.}(2024)\citenamefont {Ta{\d s}er}, \citenamefont {Do{\~g}ru}, \citenamefont {Eraslan},\ and\ \citenamefont {Ayd{\i}n}}]{taser2024conformal}%
  \BibitemOpen
  \bibfield  {author} {\bibinfo {author} {\bibfnamefont {D.}~\bibnamefont {Ta{\d s}er}}, \bibinfo {author} {\bibfnamefont {M.~U.}\ \bibnamefont {Do{\~g}ru}}, \bibinfo {author} {\bibfnamefont {E.}~\bibnamefont {Eraslan}},\ and\ \bibinfo {author} {\bibfnamefont {H.}~\bibnamefont {Ayd{\i}n}},\ }\href@noop {} {\bibfield  {journal} {\bibinfo  {journal} {Phys. Scr}\ }\textbf {\bibinfo {volume} {99}},\ \bibinfo {pages} {075049} (\bibinfo {year} {2024})}\BibitemShut {NoStop}%
\bibitem [{\citenamefont {Zee}(2010)}]{zee2010quantum}%
  \BibitemOpen
  \bibfield  {author} {\bibinfo {author} {\bibfnamefont {A.}~\bibnamefont {Zee}},\ }\href@noop {} {\emph {\bibinfo {title} {Quantum field theory in a nutshell}}},\ Vol.~\bibinfo {volume} {7}\ (\bibinfo  {publisher} {Princeton university press},\ \bibinfo {year} {2010})\BibitemShut {NoStop}%
\bibitem [{\citenamefont {Padmanabhan}(2016)}]{padmanabhan2016quantum}%
  \BibitemOpen
  \bibfield  {author} {\bibinfo {author} {\bibfnamefont {T.}~\bibnamefont {Padmanabhan}},\ }\href@noop {} {\emph {\bibinfo {title} {Quantum Field Theory: The Why, What and How}}}\ (\bibinfo  {publisher} {Springer},\ \bibinfo {year} {2016})\BibitemShut {NoStop}%
\bibitem [{\citenamefont {Magueijo}\ and\ \citenamefont {Smolin}(2004)}]{magueijo2004gravity}%
  \BibitemOpen
  \bibfield  {author} {\bibinfo {author} {\bibfnamefont {J.}~\bibnamefont {Magueijo}}\ and\ \bibinfo {author} {\bibfnamefont {L.}~\bibnamefont {Smolin}},\ }\href@noop {} {\bibfield  {journal} {\bibinfo  {journal} {Classical and Quantum Gravity}\ }\textbf {\bibinfo {volume} {21}},\ \bibinfo {pages} {1725} (\bibinfo {year} {2004})}\BibitemShut {NoStop}%
\bibitem [{\citenamefont {Chatrabhuti}\ \emph {et~al.}(2016)\citenamefont {Chatrabhuti}, \citenamefont {Yingcharoenrat},\ and\ \citenamefont {Channuie}}]{chatrabhuti2016starobinsky}%
  \BibitemOpen
  \bibfield  {author} {\bibinfo {author} {\bibfnamefont {A.}~\bibnamefont {Chatrabhuti}}, \bibinfo {author} {\bibfnamefont {V.}~\bibnamefont {Yingcharoenrat}},\ and\ \bibinfo {author} {\bibfnamefont {P.}~\bibnamefont {Channuie}},\ }\href@noop {} {\bibfield  {journal} {\bibinfo  {journal} {Physical Review D}\ }\textbf {\bibinfo {volume} {93}},\ \bibinfo {pages} {043515} (\bibinfo {year} {2016})}\BibitemShut {NoStop}%
\bibitem [{\citenamefont {Channuie}(2019)}]{channuie2019deformed}%
  \BibitemOpen
  \bibfield  {author} {\bibinfo {author} {\bibfnamefont {P.}~\bibnamefont {Channuie}},\ }\href@noop {} {\bibfield  {journal} {\bibinfo  {journal} {The European Physical Journal C}\ }\textbf {\bibinfo {volume} {79}},\ \bibinfo {pages} {508} (\bibinfo {year} {2019})}\BibitemShut {NoStop}%
\bibitem [{\citenamefont {Hendi}\ \emph {et~al.}(2016)\citenamefont {Hendi}, \citenamefont {Momennia}, \citenamefont {Panah},\ and\ \citenamefont {Faizal}}]{hendi2016nonsingular}%
  \BibitemOpen
  \bibfield  {author} {\bibinfo {author} {\bibfnamefont {S.~H.}\ \bibnamefont {Hendi}}, \bibinfo {author} {\bibfnamefont {M.}~\bibnamefont {Momennia}}, \bibinfo {author} {\bibfnamefont {B.~E.}\ \bibnamefont {Panah}},\ and\ \bibinfo {author} {\bibfnamefont {M.}~\bibnamefont {Faizal}},\ }\href@noop {} {\bibfield  {journal} {\bibinfo  {journal} {The Astrophysical Journal}\ }\textbf {\bibinfo {volume} {827}},\ \bibinfo {pages} {153} (\bibinfo {year} {2016})}\BibitemShut {NoStop}%
\bibitem [{\citenamefont {Hendi}\ \emph {et~al.}(2017)\citenamefont {Hendi}, \citenamefont {Panahiyan}, \citenamefont {Upadhyay},\ and\ \citenamefont {Eslam~Panah}}]{hendi2017charged}%
  \BibitemOpen
  \bibfield  {author} {\bibinfo {author} {\bibfnamefont {S.}~\bibnamefont {Hendi}}, \bibinfo {author} {\bibfnamefont {S.}~\bibnamefont {Panahiyan}}, \bibinfo {author} {\bibfnamefont {S.}~\bibnamefont {Upadhyay}},\ and\ \bibinfo {author} {\bibfnamefont {B.}~\bibnamefont {Eslam~Panah}},\ }\href@noop {} {\bibfield  {journal} {\bibinfo  {journal} {Physical Review D}\ }\textbf {\bibinfo {volume} {95}},\ \bibinfo {pages} {084036} (\bibinfo {year} {2017})}\BibitemShut {NoStop}%
\bibitem [{\citenamefont {Feng}\ and\ \citenamefont {Yang}(2017)}]{feng2017thermodynamic}%
  \BibitemOpen
  \bibfield  {author} {\bibinfo {author} {\bibfnamefont {Z.-W.}\ \bibnamefont {Feng}}\ and\ \bibinfo {author} {\bibfnamefont {S.-Z.}\ \bibnamefont {Yang}},\ }\href@noop {} {\bibfield  {journal} {\bibinfo  {journal} {Physics Letters B}\ }\textbf {\bibinfo {volume} {772}},\ \bibinfo {pages} {737} (\bibinfo {year} {2017})}\BibitemShut {NoStop}%
\bibitem [{\citenamefont {Hendi}\ and\ \citenamefont {Momennia}(2018)}]{hendi2018ads}%
  \BibitemOpen
  \bibfield  {author} {\bibinfo {author} {\bibfnamefont {S.~H.}\ \bibnamefont {Hendi}}\ and\ \bibinfo {author} {\bibfnamefont {M.}~\bibnamefont {Momennia}},\ }\href@noop {} {\bibfield  {journal} {\bibinfo  {journal} {Physics Letters B}\ }\textbf {\bibinfo {volume} {777}},\ \bibinfo {pages} {222} (\bibinfo {year} {2018})}\BibitemShut {NoStop}%
\bibitem [{\citenamefont {Panahiyan}\ \emph {et~al.}(2019)\citenamefont {Panahiyan}, \citenamefont {Hendi},\ and\ \citenamefont {Riazi}}]{panahiyan2019ads4}%
  \BibitemOpen
  \bibfield  {author} {\bibinfo {author} {\bibfnamefont {S.}~\bibnamefont {Panahiyan}}, \bibinfo {author} {\bibfnamefont {S.}~\bibnamefont {Hendi}},\ and\ \bibinfo {author} {\bibfnamefont {N.}~\bibnamefont {Riazi}},\ }\href@noop {} {\bibfield  {journal} {\bibinfo  {journal} {Nuclear Physics B}\ }\textbf {\bibinfo {volume} {938}},\ \bibinfo {pages} {388} (\bibinfo {year} {2019})}\BibitemShut {NoStop}%
\bibitem [{\citenamefont {Dehghani}(2018)}]{dehghani2018thermodynamics}%
  \BibitemOpen
  \bibfield  {author} {\bibinfo {author} {\bibfnamefont {M.}~\bibnamefont {Dehghani}},\ }\href@noop {} {\bibfield  {journal} {\bibinfo  {journal} {Physics Letters B}\ }\textbf {\bibinfo {volume} {777}},\ \bibinfo {pages} {351} (\bibinfo {year} {2018})}\BibitemShut {NoStop}%
\bibitem [{\citenamefont {Frassino}\ and\ \citenamefont {Panella}(2012)}]{frassino2012casimir}%
  \BibitemOpen
  \bibfield  {author} {\bibinfo {author} {\bibfnamefont {A.~M.}\ \bibnamefont {Frassino}}\ and\ \bibinfo {author} {\bibfnamefont {O.}~\bibnamefont {Panella}},\ }\href@noop {} {\bibfield  {journal} {\bibinfo  {journal} {Verhandlungen der Deutschen Physikalischen Gesellschaft}\ } (\bibinfo {year} {2012})}\BibitemShut {NoStop}%
\bibitem [{\citenamefont {Harko}\ and\ \citenamefont {Lobo}(2014)}]{harko2014generalized}%
  \BibitemOpen
  \bibfield  {author} {\bibinfo {author} {\bibfnamefont {T.}~\bibnamefont {Harko}}\ and\ \bibinfo {author} {\bibfnamefont {F.~S.}\ \bibnamefont {Lobo}},\ }\href@noop {} {\bibfield  {journal} {\bibinfo  {journal} {Galaxies}\ }\textbf {\bibinfo {volume} {2}},\ \bibinfo {pages} {410} (\bibinfo {year} {2014})}\BibitemShut {NoStop}%
\bibitem [{\citenamefont {Gibbons}\ and\ \citenamefont {Werner}(2008)}]{gibbons2008applications}%
  \BibitemOpen
  \bibfield  {author} {\bibinfo {author} {\bibfnamefont {G.}~\bibnamefont {Gibbons}}\ and\ \bibinfo {author} {\bibfnamefont {M.}~\bibnamefont {Werner}},\ }\href@noop {} {\bibfield  {journal} {\bibinfo  {journal} {Classical and Quantum Gravity}\ }\textbf {\bibinfo {volume} {25}},\ \bibinfo {pages} {235009} (\bibinfo {year} {2008})}\BibitemShut {NoStop}%
\bibitem [{\citenamefont {Kavya}\ \emph {et~al.}(2022)\citenamefont {Kavya}, \citenamefont {Venkatesha}, \citenamefont {Mandal},\ and\ \citenamefont {Sahoo}}]{kavya2022constraining}%
  \BibitemOpen
  \bibfield  {author} {\bibinfo {author} {\bibfnamefont {N.~S.}\ \bibnamefont {Kavya}}, \bibinfo {author} {\bibfnamefont {V.}~\bibnamefont {Venkatesha}}, \bibinfo {author} {\bibfnamefont {S.}~\bibnamefont {Mandal}},\ and\ \bibinfo {author} {\bibfnamefont {P.~K.}\ \bibnamefont {Sahoo}},\ }\href@noop {} {\bibfield  {journal} {\bibinfo  {journal} {Physics of the Dark Universe}\ }\textbf {\bibinfo {volume} {38}},\ \bibinfo {pages} {101126} (\bibinfo {year} {2022})}\BibitemShut {NoStop}%
\end{thebibliography}%

\end{document}